\documentclass[conference]{IEEEtran}
\IEEEoverridecommandlockouts

\usepackage{algorithmic}
\usepackage{textcomp}
\usepackage{xcolor}
\usepackage{amsmath}
\usepackage{graphicx}
\usepackage{siunitx}
\usepackage{multirow}
\usepackage{booktabs}
\usepackage{cleveref}
\usepackage{comment}

\def\BibTeX{{\rm B\kern-.05em{\sc i\kern-.025em b}\kern-.08em
    T\kern-.1667em\lower.7ex\hbox{E}\kern-.125emX}}

\usepackage[style=ieee,backend=biber, url=false, doi=false, isbn=false]{biblatex}

\Crefname{figure}{Fig.}{Fig.}

\begin{document}

\title{
Equity-aware Load Shedding Optimization
}

\author{\IEEEauthorblockN{Xin Fang}
\IEEEauthorblockA{\textit{Department of Electrical \& Computer Engineering} \\
\textit{Mississippi State University}\\
Starkville, MS, USA \\
\ xfang@ece.msstate.edu}
\and
\IEEEauthorblockN{Wenbo Wang, Fei Ding}
\IEEEauthorblockA{\textit{Energy System Integration}\\
\textit{National Renewable Energy Laboratory}\\
Golden, CO, USA\\
\{wenbo.wang, fei.ding\}@nrel.gov}
}
\maketitle

\begin{abstract}
Load shedding is usually the last resort to balance generation and demand to maintain stable operation of the electric grid after major disturbances. Current load-shedding optimization practices focus mainly on the physical optimality of the network power flow. This might lead to an uneven allocation of load curtailment, disadvantaging some loads more than others. Addressing this oversight, this paper introduces an innovative equity-aware load-shedding optimization model that emphasizes a fair allocation of load curtailment across the network. By proposing a novel equity indicator for load shedding and integrating it into an ACOPF-based optimization framework, we offer grid operators a more balanced and equitable load shedding strategy. Case studies highlight the importance of equity considerations in determining optimal load curtailment between buses.
\end{abstract}


\section{Introduction}
The reliability and security of power systems have always been of utmost importance to utilities and system operators to improve the reliability of the electricity supply to customers. To ensure these criteria, one of the last resort control measures after major contingencies is load shedding \cite{grewal_optimization_1998}, which helps maintain the balance of the supply and demand of the system by curtailing the power load in certain areas. As power systems evolve with the integration of renewable resources, smart grids, and advanced control systems, it is imperative to reevaluate conventional load-shedding schemes. A major concern that emerges is how to ensure that load shedding is implemented equitably, improving fairness among consumers to ensure that customers at different locations have an equitable risk of load curtailment \cite{noauthor_energy_2020,jenkins_energy_2016}. 

In designing an effective load-shedding strategy, three questions must be addressed: 1) when should load-shedding be initiated; 2) where should the load be shed; 3) what method should be used for load-shedding and how much load needs to be shed \cite{zhang_deep_2023}. Traditionally, load-shedding decisions are made based on technical priorities such as stability, frequency control, power losses, voltage management, and importance of the load \cite{elyasichamazkoti_optimal_2022}. Although these are undeniably vital for system operation, an overly narrow focus on these factors can unintentionally lead to inequitable service interruptions among customers, exacerbating socioeconomic disparities. Customers located in some areas might face an unproportionate risk of load curtailment due to the limitations of the networks (e.g., network losses and congestion) and generation resources. For example, power-loss minimization-based load-shedding algorithms will be prone to curtail loads at remote nodes compared to near nodes to reduce power losses on the lines. Although optimal from a physical perspective, these load-shedding results lack equity consideration and lead to unfair electricity interruption within specific regions. Such disparities may further lead to trust erosion between consumers and utilities, necessitating a relook at how we balance technical requirements with equity considerations.

Although there is extensive research on optimal load-shedding strategies that focus on minimizing the curtailed load or enhancing system stability \cite{fitri_priority-considered_2023,liu_robust_2022}, the lens of equity has not been widely addressed. Some studies have pointed out the social implications of load shedding \cite{noauthor_community_nodate,wang_strategy_2019}, but systematic methodologies that incorporate equity as an integral part of load shedding optimization and other grid operation frameworks are still unexplored \cite{fang_distributed_2023}. The lack of a transparent load-shedding equity indicator further complicates this effort. The Gini coefficient, widely used in socioeconomic evaluations, is a measure of statistical dispersion intended to represent income inequality \cite{noauthor_gini_2023}. This concept offers promising insights that can be translated into the context of load-shedding optimization in power systems.

This paper aims to bridge the aforementioned gaps in the literature and contribute to a more equitable load-shedding framework. Our primary contributions can be summarized as follows.
\begin{enumerate}
    \item \textbf{Grid Gini Coefficient for Outage Risk Equity}: We introduce a novel grid Gini coefficient, adapted from its economic counterpart, as an outage equity indicator. This metric serves as a robust tool for evaluating the equity of load shedding across different regions, providing a standardized measure to capture disparities.
    \item \textbf{Equity-Aware Load Shedding Optimization Model}: Recognizing the need to balance technical constraints with equity considerations harmoniously, we present an equity-aware load-shedding optimization model. This model aims to guide load-shedding strategies to ensure an equitable distribution of service interruptions among various load buses.
    \item \textbf{Interaction between Equity Constraints and Physical Limitations}: We conduct comprehensive case studies to support the proposed methodologies. These explore the impact of incorporating equity constraints on load-shedding strategies, especially in the context of the generation trip N-1 contingency.
\end{enumerate}

The remainder of this paper is organized as follows. Section II proposes a novel grid Gini coefficient for the load-shedding equity indicator to represent the disparity between load-shedding buses. Section III formulates the equity-constrained load-shedding optimization model based on the alternate current optimal power flow (ACOPF), including an equity constraint. Section IV performs the case studies in a revised IEEE 14-bus system. Section V concludes the paper.

\section{Equity Indicator of Load Shedding}
\begin{figure}
    \centering
    \vspace{-1em}
    \includegraphics[width=0.9\linewidth]{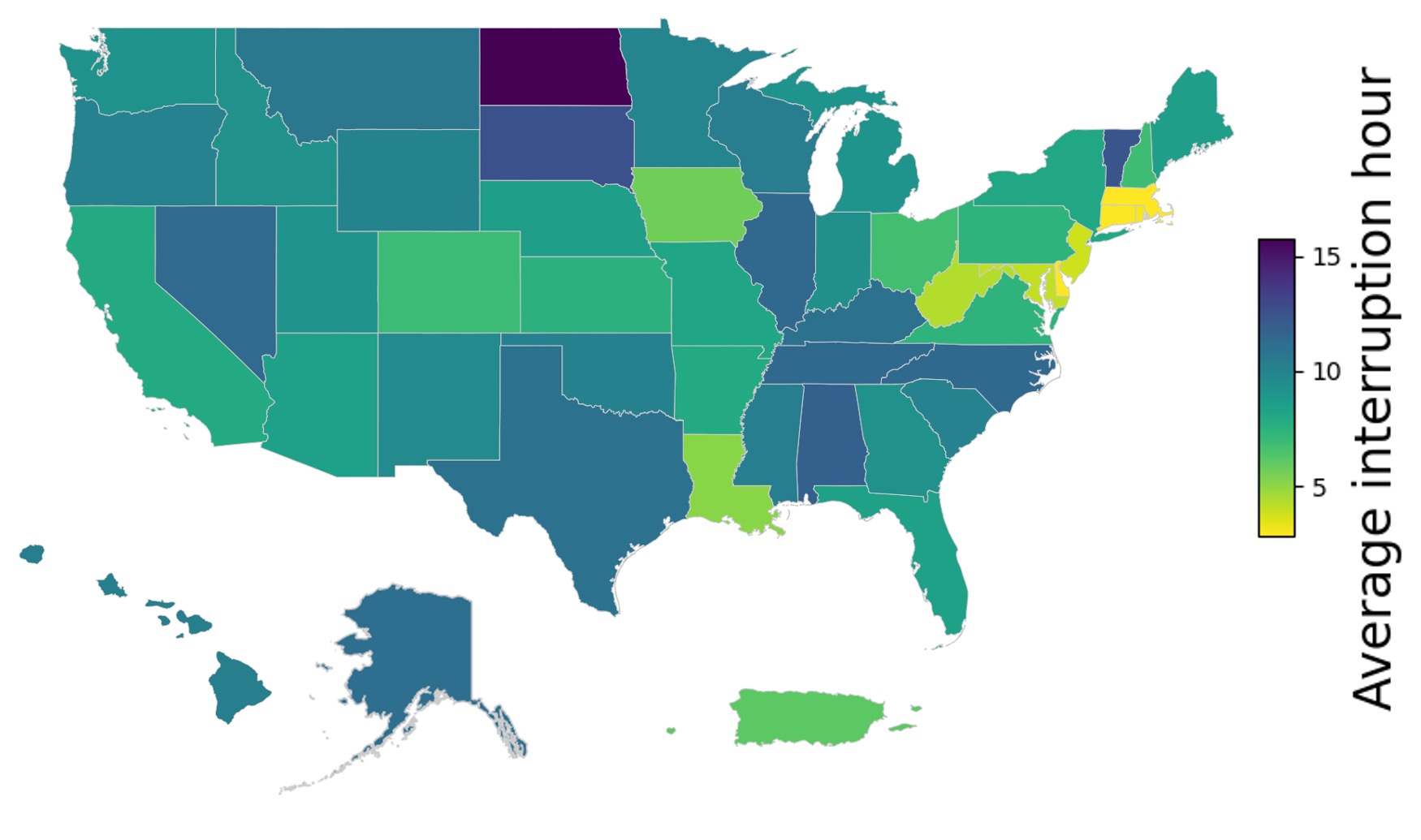}
    \vspace{-2.5em}
    \caption{Electricity outage average interruption hours in different regions}
    \label{fig:outagedistribution}
\end{figure}

Historically, in the load-shedding optimization scheme, the total curtailed load has been a prevalent metric to assess the risk of loss of electricity within the grid \cite{ramezani_determination_2009, ortega-vazquez_estimating_2009,tomasson_generation_2018,liu_quantifying_2012}. Although these metrics offer a broad overview of the system, they tend to overlook the intricate geographic inequalities. In fact, power outages are far from random; they are correlated with geographical locations, geographical disparities are evident, as shown in Fig. \ref{fig:outagedistribution}. For example, states like \textbf{North Dakota} faced interruptions that lasted several times longer than other states, as shown in Fig. \ref{fig:outagedistribution}. The future generation mix, represented by geographically dependent RES and DER, will only elevate these disparities. Evidenced by outage reports \cite{noauthor_june-august_nodate,noauthor_mayjune_nodate}, inverter-based RES and DER exhibit sensitivity to system frequency and voltage fluctuations. With regions showing varied reactions to frequency and voltage disturbances after contingencies, there is a pressing need to improve outage metrics based on the total curtailed load. To enhance these metrics, a grid Gini coefficient is proposed for the outage equity indicator to encompass the geographical intricacies of outage risk and guide a more equitable allocation of load curtailment.

To incorporate equity considerations into load-shedding optimization models after N-1 contingencies, we first design the outage equity indicator of power outages on the grid. To transparently evaluate the equity of load shedding among various buses with different socioeconomic backgrounds, a novel grid Gini coefficient is proposed for the outage equity indicator ($GGC_{OEI}$), supported by locational load shedding metrics. The outage/shedding risk index for bus outages $ORI$ and the grid Gini coefficient for the load shedding equity indicator ($GGC_{OEI}$) are shown below.

\begin{equation}
ORI_i= \frac{P_{LS,i}}{P_{Li}}
\end{equation}
\begin{equation}
GGC_{OEI} = \frac{\sum_{i=1}^{n} \sum_{j=1}^{n} |ORI_i - ORI_j|}{2n\times(n-1)\times ORI_{mean}}
\label{EquityIndicator}
\end{equation}
\begin{equation}
ORI_{mean} = \frac{1}{n} \sum_{j=1}^{n}ORI_i
\end{equation}

\noindent where $ORI_i$ is the outage risk indicator for socioeconomic subregion \textit{i}; ${P_{Li}}$ is the demand for socioeconomic subregion \textit{i}; $P_{LS,i}$ is the load shedding of the subregions \textit{i}; $ORI_{mean}$ is the mean ORI of all subregions; $n$ is the number of subregions. In the socioeconomic evaluation, a high Gini coefficient represents an inequitable distribution of wealth within the region \cite{noauthor_gini_2023}. Similarly, a high $GGC_{OEI}$ represents an inequitable distribution of load-shedding within the network.

\section{Equity-constrained Load Shedding Optimization}

To minimize the curtailed load, an optimization-based model is usually used. The formulation of the ACOPF-based load-shedding model is shown below. In addition to traditional generation power limitations, line power flow, and voltage angle/magnitude constraints, a new load-shedding equity constraint is based on the equity indication proposed in the previous section. The cost function is the traditional generation cost plus the penalty cost of load-shedding. In this model, each bus is assumed to be a subregion for simplicity.
\vspace{0.3em}
\begin{equation}
   min \sum_{g\in gen} f(P_{g})+\sum_{Li\in Load} C_{LS}*P_{LS,Li} 
\end{equation}

Subject to:

Load shedding equity constraint:
\vspace{0.3em}
\begin{equation}
GGC_{OEI}  \leq \beta
\label{eq:equityconstraint}
\end{equation}

Nodal active/reactive power equations:
\begin{multline}
  P_{Gi}-(P_{Li}-P_{LS,i})\\
  =V_{i}\sum_{j=1}^{n}V_{j}(G_{i,j}cos(\theta_{i}-\theta_{j})+B_{ij}cos(\theta_{i}-\theta_{j})) 
\end{multline}
\vspace{-1em}
\begin{multline}
  Q_{Gi}-(Q_{Li}-Q_{LS,i})\\
  =V_{i}\sum_{j=1}^{n}V_{j}(G_{i,j}sin(\theta_{i}-\theta_{j})-B_{ij}sin(\theta_{i}-\theta_{j})) 
\end{multline} 

Generation operational power constraints:
\vspace{0.3em}
\begin{equation}
  P_{min,g} \leq P_{g}\leq P_{max,g}, g\in gen
\end{equation} 
\begin{equation}
  Q_{min,g} \leq Q_{g}\leq Q_{max,g}, g\in gen
\end{equation} 

Network constraints:
\vspace{0.3em}
\begin{equation}
    V_{min,i}\leq V_{i}\leq V_{max,i}, i\in bus
\end{equation} 
\begin{equation}
    L_{l}\leq L_{max,l}, l\in line
\end{equation}

\noindent where $f(P_{g})$ is the cost function of the conventional generation resources; $C_{LS}$ is the penalty price for load curtailment; $P_{LS,i}$ is the real power curtailment at bus $i$; $Q_{LS,i}$ is the reactive power curtailment at bus $i$; $\beta$ is the limit for the $GGC_{OEI}$ in the load-shedding; $P_{Gi}$ and $Q_{Gi}$ are the total active/reactive power generation at Bus $i$; $P_{Li}$ and $Q_{Li}$ are the bus $i$ active and reactive power demand; $P_{g}$ and $Q_{g}$ represents the active and reactive power output of the g-th generator; $P_{min,g}$ and $P_{max,g}$ are the minimum and maximum active generation of the g-th generator; $Q_{min,g}$ and $Q_{max,g}$ are the minimum and maximum reactive generation of the g-th generator; $\theta_{i}$ is the voltage angle of bus $i$; $V_{i}$ is the voltage magnitude of Bus $i$; $V_{min,i}$ and $V_{max,i}$ are the upper and lower bounds of bus voltage; $L_{l}$ is the branch power flow; $L_{max,l}$ is the branch flow limit. Note that a smaller $GGC_{OEI}$ represents a more equitable load-shedding while a large $GGC_{OEI}$ means that the load is concentratedly curtailed on several buses. 

Combining the $GGC_{OEI}$ definition Eq. \ref{EquityIndicator} and the equity constraint $GGC_{OEI} \leq \beta$, it can be formulated as follows. 

\begin{equation}
\sum_{i=1}^{n} \sum_{j=1}^{n} |ORI_i - ORI_j| \leq \beta \cdot 2n\times(n-1)\times ORI_{mean}
\end{equation}

\noindent where the absolute term $|ORI_i - ORI_j|$ can be further linearized by introducing auxiliary variables $z_{i,j}^+$ and $z_{i,j}^-$. 
Then, this equity constraint will be converted to linear constraints, and the whole model will be solved by the nonlinear solver CONOPT \cite{noauthor_conopt_nodate}. 

\section{Case Studies}
The IEEE 14-bus system is used to demonstrate the impact of the proposed equity constraints on the optimization of load shedding. The mathematical optimization model of the proposed equity-aware load shedding is programmed in the general algebraic modeling system (GAMS) with a nonlinear solver of CONOPT \cite{noauthor_gams_nodate,noauthor_conopt_nodate}. The simulation is performed on a Macbook Pro with an M2 CPU and 8GB RAM.

\vspace{-0.4em}
\subsection{Parameters of the Testing System}
\begin{figure}[h]
    \centering
    \includegraphics[width=0.7\linewidth]{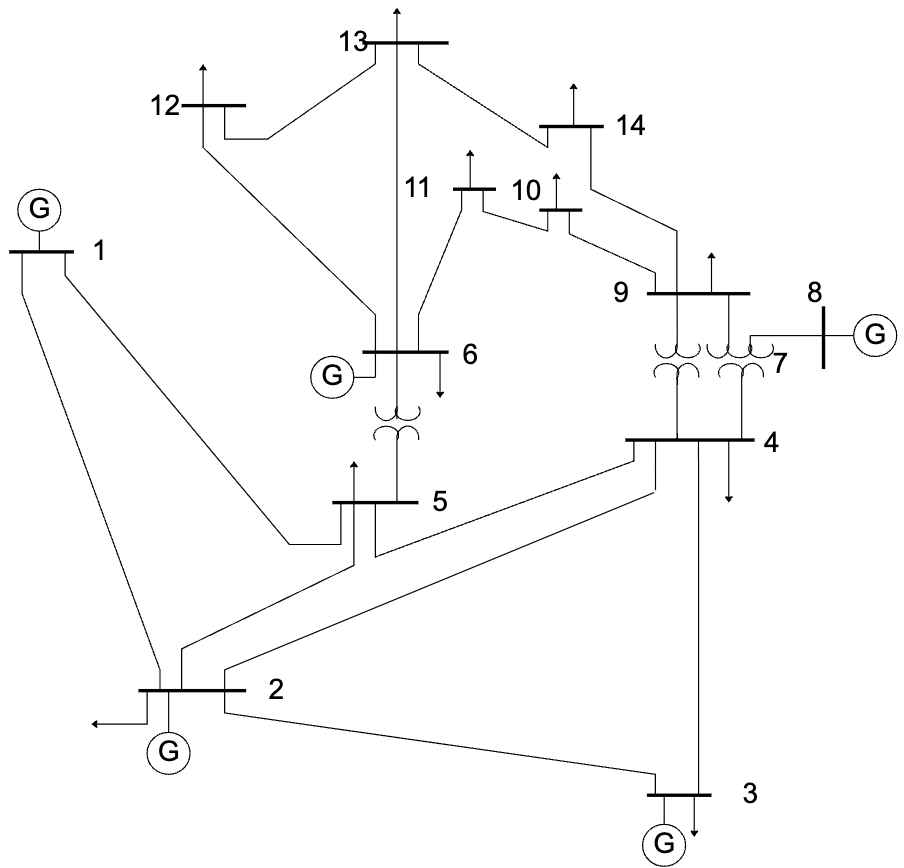}
    \caption{IEEE 14-bus test system network}
    \label{fig:14bus}
\end{figure}

The real power demand on each load bus is increased by 100\% to stress the system and demonstrate the impact of generation loss and load-shedding. The network of this testing system is depicted in Fig. \ref{fig:14bus}. The demand for buses is shown in Table \ref{tab:bus_load}.  The generator parameters are listed in Table \ref{tab:Gen_info}. All other parameters, including line parameters and generator cost coefficients, are the same as in the Matpower 14-bus test system. The price of the penalty for load-shedding $C_{LS}$ is \$ 500000/MWh in the case study. In load-shedding optimization, the largest generator with a capacity of 332MW at Bus 1 is tripped to mimic the system's severest N-1 generation contingency.

\begin{table}[h]
    \centering
\caption{Bus real power load}
\label{tab:bus_load}
    \begin{tabular}{|c|c|l|} \hline 
         Bus& Pd (MW) &Qd (MVar)\\ \hline 
         2& 43.4 &12.7\\ \hline 
         3& 188.4 &19\\ \hline 
         4& 95.6 &-3.9\\ \hline 
         5& 15.2 &1.6\\ \hline 
         6& 22.4 &7.5\\ \hline 
         9& 59 &16.6\\ \hline 
         10& 18 &5.8\\ \hline 
         11& 7 &1.8\\ \hline 
         12& 12.2 &1.6\\ \hline 
 13&27 &5.8\\ \hline 
 14&29.8 &5\\ \hline
    \end{tabular}  
\end{table}

\begin{table}[h]
    \centering
\caption{Generator parameters in IEEE 14-bus system}
\label{tab:Gen_info}
    \begin{tabular}{|c|c|c|l|} \hline 
         Generator&  Bus& Capacity (MW)&Status\\ \hline 
         1&  1&  332&0\\ \hline 
         2&  2&  140&1\\ \hline 
         3&  3&  100&1\\ \hline 
         4&  6&  100&1\\ \hline 
         5&  8&  100&1\\ \hline
    \end{tabular} 
\end{table}

\vspace{-0.4em}
\subsection{Generation Dispatch before N-1 Generation Loss}
Before the G1 tripping, Table \ref{tab:Gen-dispatch} lists the system's generation dispatch. It is clear that the largest generator, G1, has the highest power output of 214.9 MW. The total generation cost before the contingency is \$18877.4.  All units have real power outputs to service the load of the system.

\begin{table}
    \centering
\caption{Generation dispatch before G1 loss contingency}
\label{tab:Gen-dispatch}
    \begin{tabular}{|c|c|} \hline 
         Generator& Real power dispatch (MW)\\ \hline 
         G1& 214.9\\ \hline 
         G2& 41.2\\ \hline 
         G3& 100
\\ \hline 
         G4& 78.1\\ \hline 
         G5& 97.6\\ \hline
    \end{tabular}
    
\end{table}

\vspace{-0.3em}
\subsection{Implication of Load Shedding Equity}
This subsection investigates the implications of the proposed equity metrics and the constraints on load-shedding strategies. After the largest generator in Bus 1 is tripped, the total generation capacity becomes 440 MW, as shown in Table \ref{tab:Gen_info}, which is less than the total real load capacity of 518 MW in Table \ref{fig:14bus}. Therefore, there will be load curtailment (at least 78 MW load curtailment to balance generation and load). The equity-aware load-shedding optimization model in Section III is used to optimize load curtailment. The value of $\beta$ varies from 0.05 to 1 to investigate the impact of this equity constraint. 

\begin{figure}
    \centering
    \vspace{-2em}
    \includegraphics[width=0.9\linewidth]{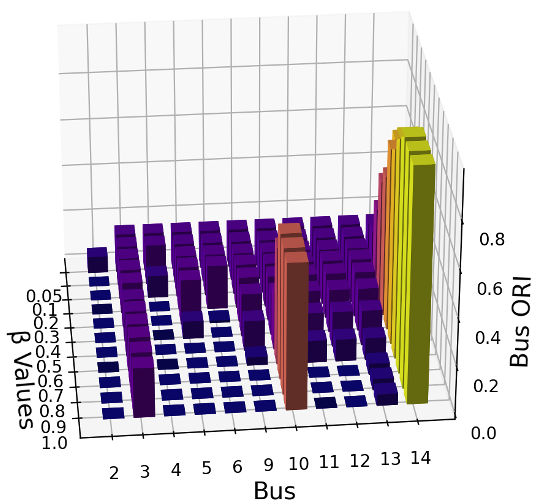}
    \vspace{-1em}
    \caption{Bus ORI under different values of $\beta$}
    \label{fig:BusORI}
\end{figure}

After the load shedding under different values of $\beta$, the bus ORI is depicted in Fig. \ref{fig:BusORI}. It is evident that when $\beta$ is high, such as greater than 0.8, most of the load curtailment will be on buses 14, 10, and 3. The load on Bus 14 has a disproportionately high risk of load curtailment during the load-shedding practice due to network constraints and losses. Considering only the physical network constraints and load shedding cost can perpetuate this load shedding disparity and lead to unfair shedding strategies. When the value of $\beta$ decreases, the loads on other buses start to be curtailed. From the $GGC_{OEI}$ formulation in Eq. \ref{EquityIndicator} and the equity constraint in Eq. \ref{eq:equityconstraint}, a small value of $\beta$ will enforce a strict equity requirement. Therefore, the load curtailment will be distributed more evenly with a small value of $\beta$. In other words, with this constraint added and an appropriate value of $\beta$, the load can be cut more equitably. In this system, when the value of $\beta$ is smaller than 0.3, the load curtailment can be distributed on most buses. With more buses' loads curtailed, the curtailment on each bus reduces. This can reduce the risk of load shedding on all buses fairly. 

\begin{figure}
    \centering
    \vspace{-1em}
    \includegraphics[width=0.9\linewidth]{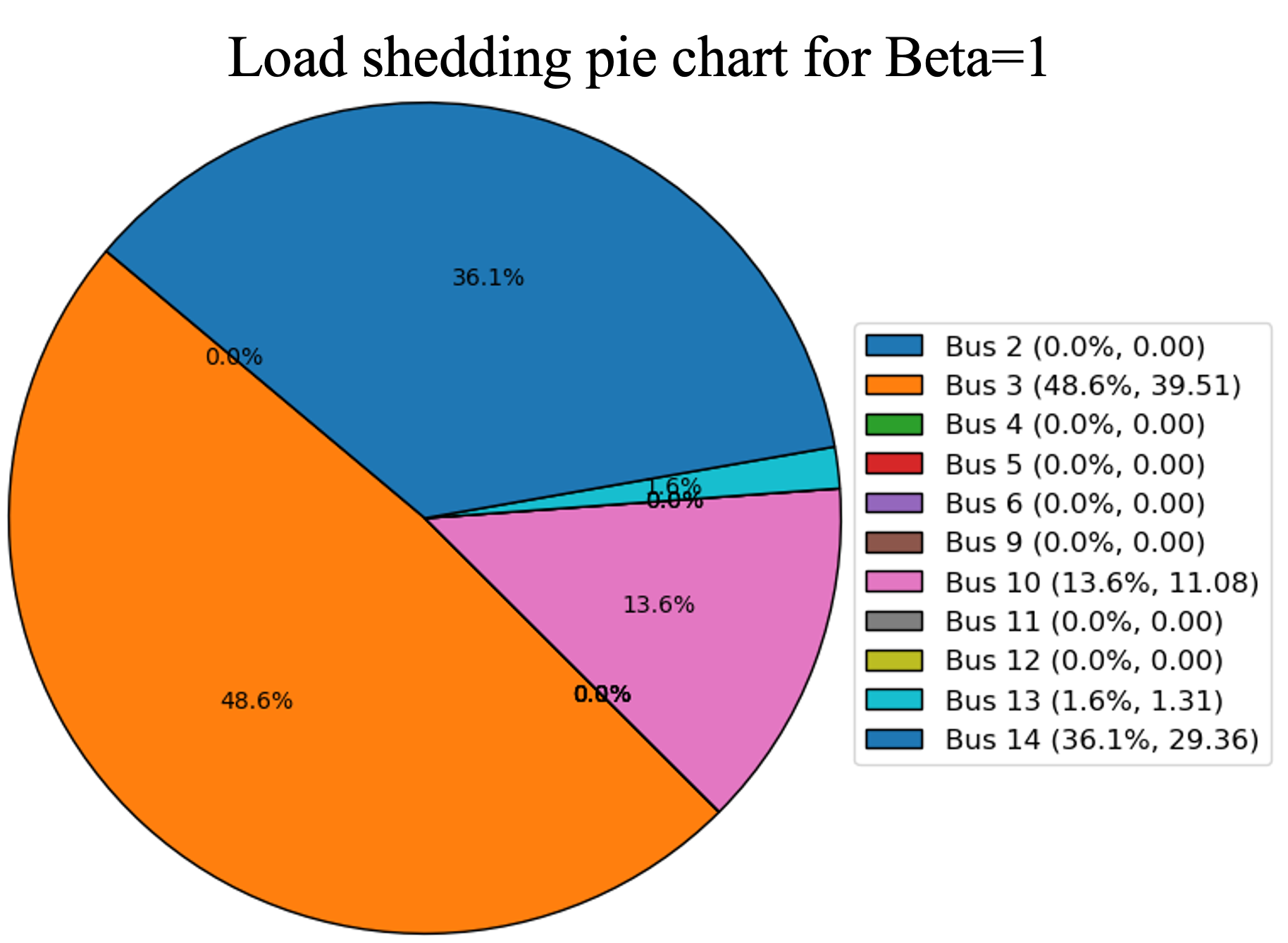}
    \vspace{-1em}
    \caption{Load shedding on buses with $\beta$=1}
    \label{fig:loadshedding_beta_1}
\end{figure}

\begin{figure}
    \centering
    \vspace{-1em}
    \includegraphics[width=0.9\linewidth]{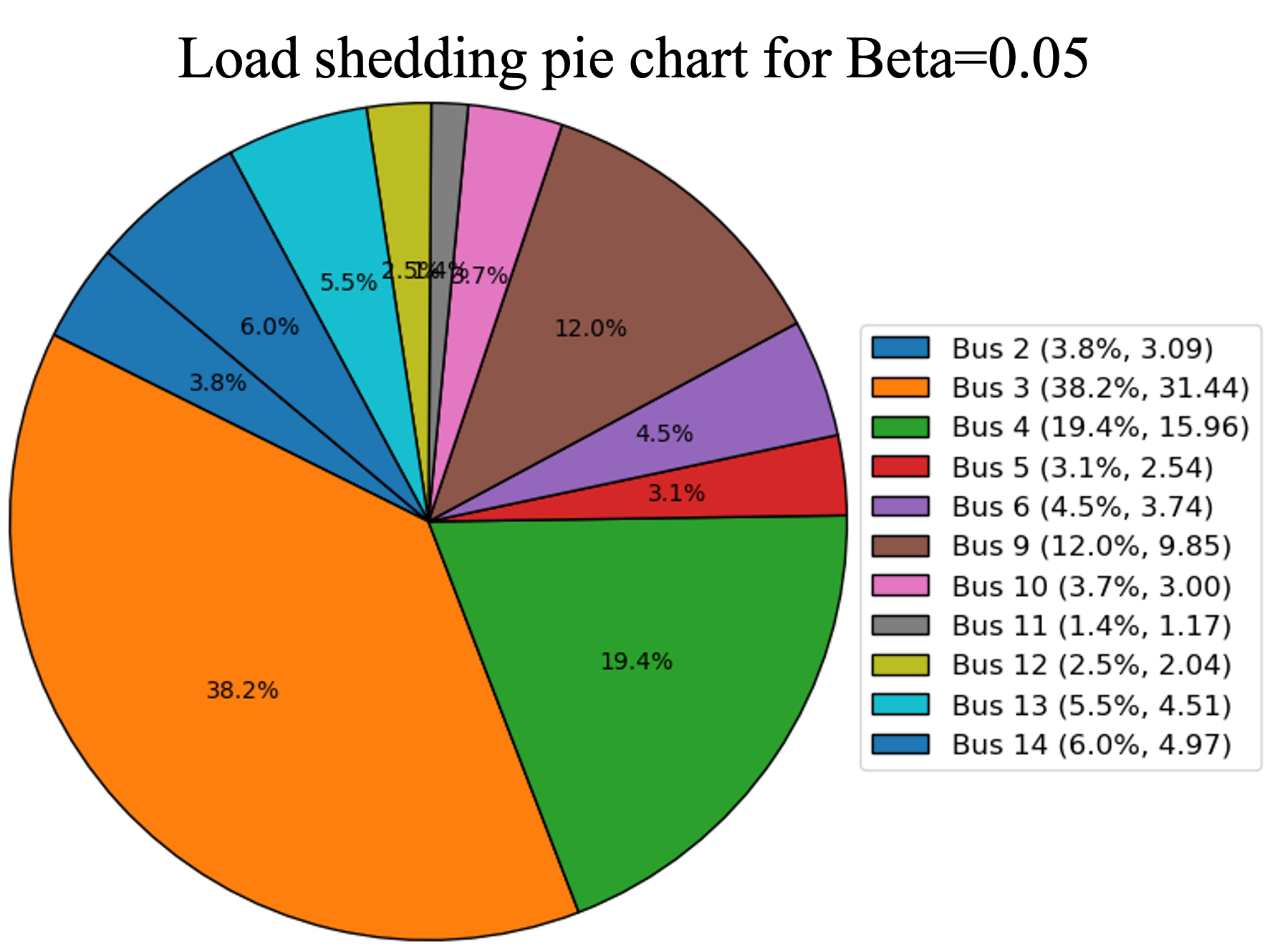}
    \vspace{-1em}
    \caption{Load shedding on buses with $\beta$=0.05}
    \vspace{-1em}
    \label{fig:loadshedding_beta_0.05}
\end{figure}

When $\beta$ is 1 and 0.05, the load curtailments on different buses are shown in Fig. \ref{fig:loadshedding_beta_1} and Fig. \ref{fig:loadshedding_beta_0.05}. Comparing these figures to Fig. \ref{fig:BusORI}, it can be found that the highest load shedding risk does not necessarily occur in the buses with the highest load shedding. For example, when $\beta$ is 1, Bus 3 has the largest load shedding shown in Fig. \ref{fig:loadshedding_beta_1}. But in Fig. \ref{fig:BusORI}, its $ORI$ is not the highest. Table \ref{tab:bus_load} shows that Bus 3 has the largest active power demand. Therefore, even with a large load shedding, its shedding risk  $ORI$ can be moderate. Bus 14 has a large load shedding leading to the highest $ORI$, with almost all of the load being shed because Bus 14 has a much smaller load than Bus 3.

Compared Fig. \ref{fig:loadshedding_beta_1} to Fig. \ref{fig:loadshedding_beta_0.05}, it can be observed that with a smaller $\beta$, more buses will have load shedding with smaller curtailment amounts. Therefore, although the total amount of shedding is slightly increased, the risk of shedding on each bus is moderate. On the contrary, without this equity constraint or with a very large $\beta$, the load shedding concentrates on Bus 3, Bus 10, and Bus 14, while the other buses do not have any shedding. This leads to unfair shedding. With a small $\beta$, equity-aware load shedding can obtain a more equitable shedding strategy that disperses the shedding to more buses with a low risk of shedding for each bus, as shown in Fig. \ref{fig:BusORI}.

\subsection{Impacts of Load Shedding Equity on Costs and $GGC_{OEI}$}

\begin{figure}
    \centering
    \includegraphics[width=0.9\linewidth]{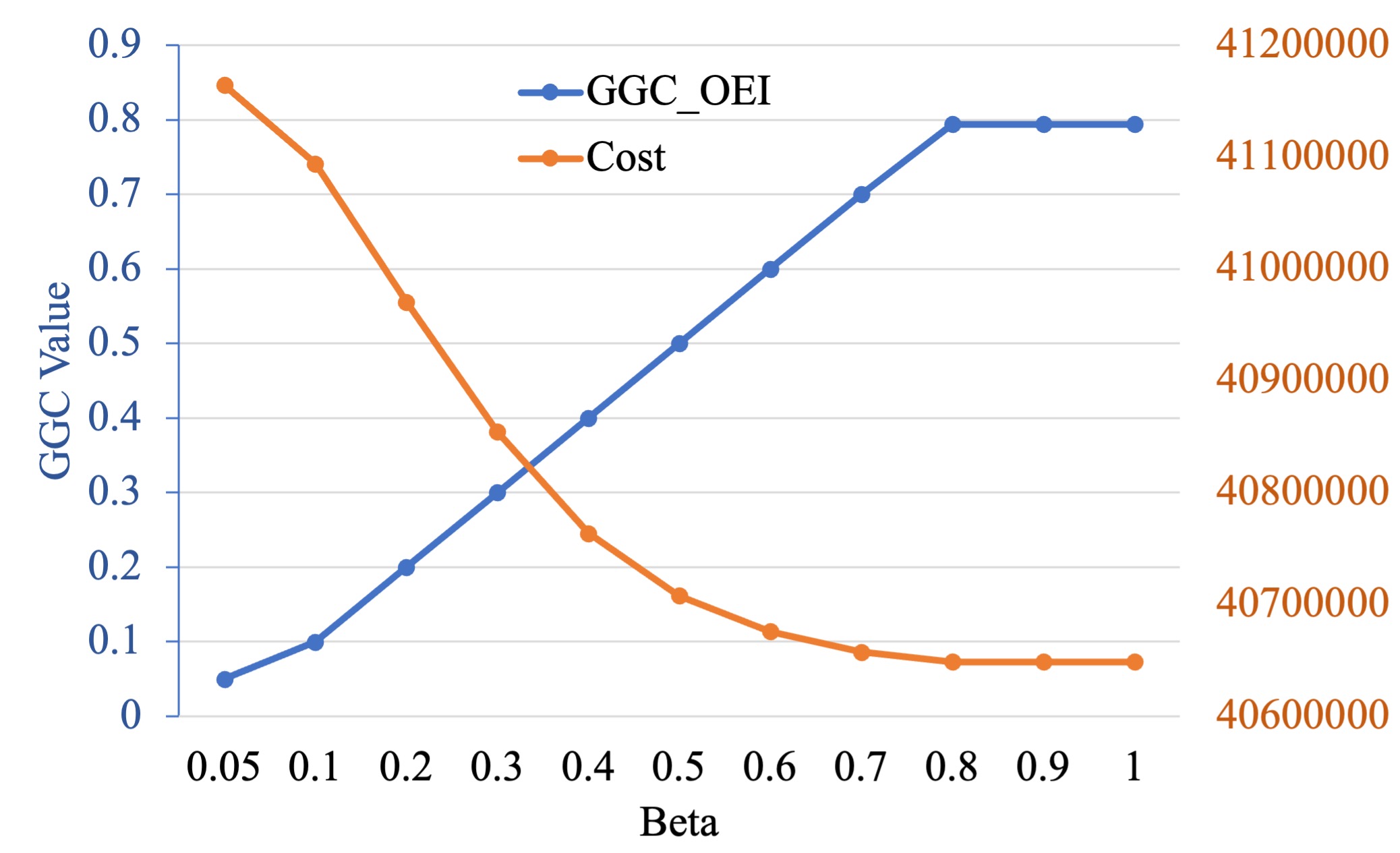}
    \vspace{-1em}
    \caption{$GGC_{OEI}$ and the total cost with different values of $\beta$}
    \label{fig:GGC_OEI}
\end{figure}

\begin{figure}
    \centering
    \includegraphics[width=0.9\linewidth]{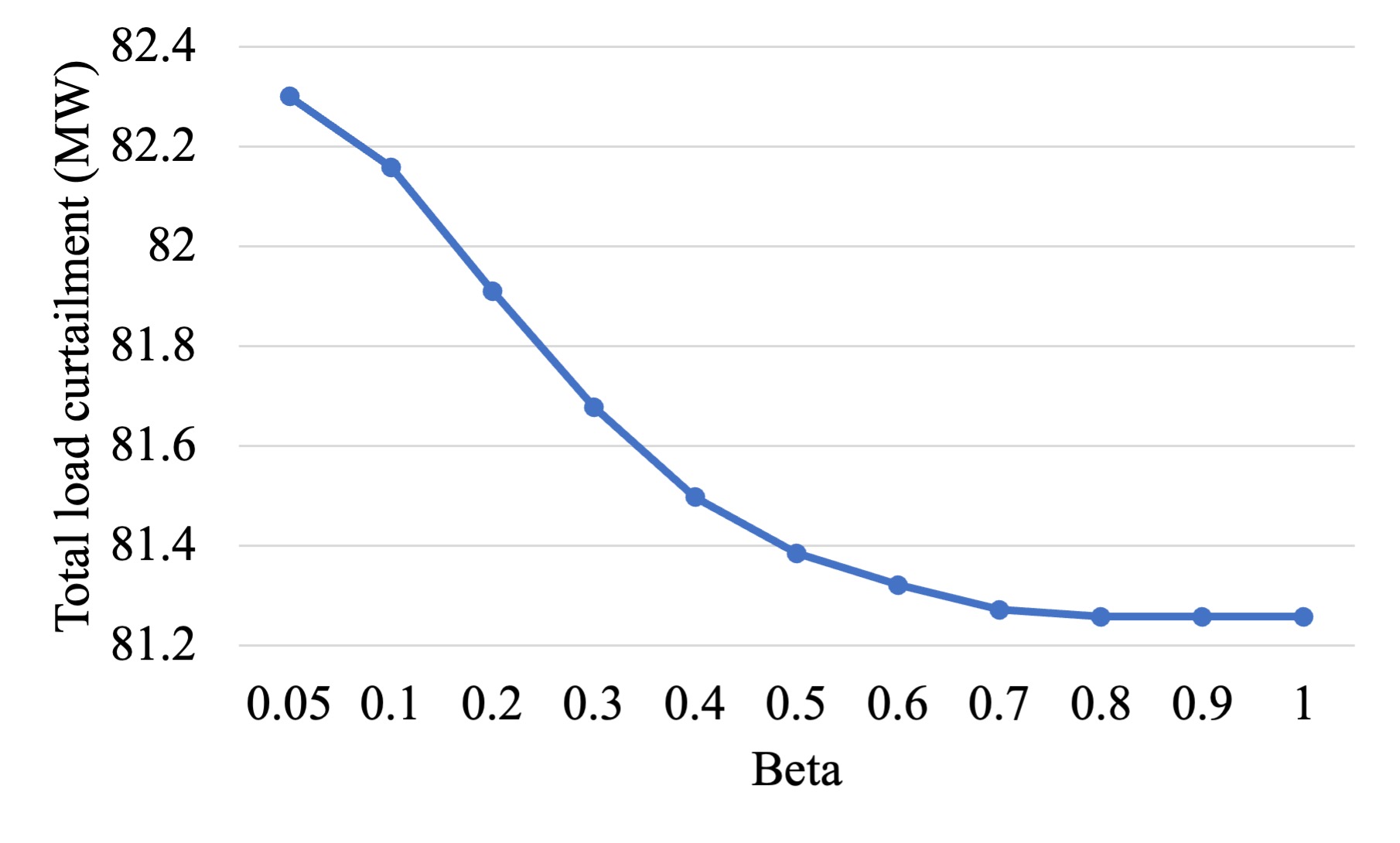}
    \vspace{-2em}
    \caption{Total load curtailment amount under different values of $\beta$}
    \vspace{-1em}
    \label{fig:loadcurtailment}
\end{figure}

This subsection analyzes the total cost of the system and the equity indicator $GGC_{OEI}$ after load shedding. The cost and $GGC_{OEI}$ under different values of $\beta$ are shown in Fig. \ref{fig:GGC_OEI}. The total curtailed load under different values of $\beta$ is shown in Fig. \ref{fig:loadcurtailment}.

Intuitively, the cost of curtailment increases with the value of $\beta$ decreases. When $\beta$ is reduced from 1 to 0.05, the total cost increases by 1.27\% from \$40648767 to \$41164191. Taking into account that the penalty price of load curtailment $C_{LS}$ is \$ 500000/MWh, the load curtailment increases by 1.03 MW when $\beta$ reduces from 1 to 0.05, also shown in Fig. \ref{fig:loadcurtailment}.

In contrast, the system's load-shedding equity indicator $GGC_{OEI}$ reduces with $\beta$. This means that the small value of $\beta$ ensures that load-shedding is more equitable among various buses. Another interesting observation is that the maximum $GGC_{OEI}$ is close to 0.8 in this system. When the value of $\beta$ is greater than 0.8, the equity constraint will not be binding. This means that the equity constraint is meaningless when choosing a large value of $\beta$. Operators should appropriately choose this $\beta$ value according to the operating conditions and network parameters. 

Figs. \ref{fig:GGC_OEI} and \ref{fig:loadcurtailment} demonstrate that improved load-shedding equity comes with a cost increase with a slightly greater load curtailment. In a real operation, the operators should also choose an appropriate $\beta$ value to trade off equity and cost in load-shedding optimization. These figures also demonstrate that the load-curtailment penalty cost will be the dominant factor in the total cost, with the generation cost becoming trivial if the penalty price is high. 

\section{Conclusion}
In this paper, we proposed an equity-aware load-shedding optimization model that incorporates equity consideration in the management of power systems. By introducing an innovative grid Gini coefficient as a load-shedding equity indicator, we introduce a clear representation of the load-curtailment disparities between buses. Case studies demonstrate that: Without equity in load shedding, specific buses face an exponentially increased risk after N-1 generation loss contingencies. Incorporating equity ensures a balanced and just curtailment across all buses, promoting a more resilient and unified grid. Although this equitable approach may marginally increase the total load curtailed, its true merit lies in its redistribution: shedding is dispersed, diluting risk, ensuring no single bus is overly affected. This approach contributes to a potential future load shedding scheme where fairness and efficiency coalesce, moving us towards a more reliable and fair power system for all.

\section*{Acknowledgments}
This work was authored in part by the National Renewable Energy Laboratory, operated by Alliance for Sustainable Energy, LLC, for the U.S. Department of Energy (DOE) under Contract No. DE-AC36-08GO28308. Funding is provided by the U.S. Department of Energy Office of Electricity Advanced Grid Research and Development program. The U.S. Government retains, and the publisher, by accepting the article for publication, acknowledges that the U.S. Government retains a nonexclusive, paid-up, irrevocable, worldwide license to publish or reproduce the published form of this work or allow others to do so, for the U.S. Government purposes. The views expressed in the article do not necessarily represent the views of the DOE or the U.S. government. 

\printbibliography
\end{document}